\documentclass[
twocolumn,
prl,letterpaper,floatfix,showpacs,superscriptaddress
]{revtex4}

\usepackage{times}
\usepackage{amsmath}
\usepackage{amsfonts}
\usepackage{amssymb}
\usepackage{bm}
\usepackage{natbib}
\usepackage{graphicx}
\usepackage{epstopdf}
\usepackage{enumerate}
\usepackage{color}

\graphicspath{{./figs/}{./}}

\begin{document}

\title{DNA Confinement drives uncoating of the HIV Virus}

\author{Ioulia Rouzina}
\affiliation{Department of Biochemistry, Molecular Biology and Biophysics, University of Minnesota, Minneapolis, MN}

\author{R.F. Bruinsma}
\affiliation{Department of Physics and Astronomy, University of California, Los Angeles, CA 90095, USA}
\affiliation{Department of Chemistry and Biochemistry, University of California, Los Angeles, CA 90095, USA}

\date{\today}

\begin{abstract}
We present a model for the uncoating of the HIV virus driven by forces exerted on the protein shell of HIV generated by DNA confinement. 
\end{abstract}

\pacs{}

\maketitle

\section{Introduction}

The confinement of double stranded (ds) viral DNA molecules has long attracted the interest of physicists. The bacteriophage viruses are the classical example of viruses where large pressures are produced by viral DNA confinement (in the range of tens of atmospheres)\cite{Kindt}. The protein shell - the \textit{capsid} - has to be very strong so as to withstand these enormous pressures. 

\begin{figure}
\centering
\includegraphics[width=0.35\textwidth]{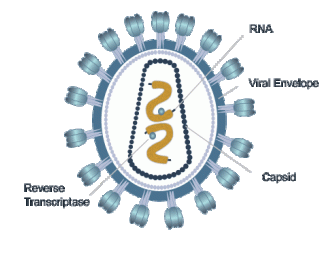}\\
\caption{\label{cartoon} Schematic cross section of a mature HIV virus. The single-stranded RNA molecule is indicated as a single helix. The capsid - the inner shell - contains, apart from the viral genome, the enzymes Integrase and Reverse Transcriptase. (credit: Ders Notlari)}
\end{figure}

In this article we propose that DNA confinement forces play a very important role in a very different class of viruses, namely the \textit{retroviruses}, such as HIV.  Figure~\ref{cartoon} shows a schematic cross section of a mature HIV virus. Maturation of an HIV virus starts from an immature spherical shell of about $10^3$ Gag ``polyproteins" that envelops both the genome molecules as well certain enzymes. This Gag protein has cationic domains that are spliced of during maturation, known as ``NCp7'' fragments that aggregate with the anionic RNA genome molecules through electrostatic interactions. Other fragments of the Gag polyprotein (``CA" fragments) form the inner capsid while the remainder of the Gag protein become part of the outer shell of the mature virus (see Fig~\ref{cartoon}). Upon infection, this outer shell fuses with the membrane of the infected cell and the mature conical capsid containing the viral RNA gets released into the host cytoplasm.

Retroviruses differ in crucial aspects from phages in terms of genome confinement. First, the genome of retroviruses is composed of two single stranded (ss) RNA molecules, each of about $10^4$ bases (10 kilobase or kb). Single stranded viral RNA molecules of 2 kb form branched structures in solution stabilized by short, complementary-paired sequences. The typical size of these structures is about 50 nm \cite{Gopal} and a 10 kb genome would be expected to be about twice larger \footnote{Because the size of an annealed branched polymer scales with the number $N$ of monomers as $N^{1/2}$.}.  The HIV RNA genome molecules are confined inside the conical capsid that has a maximum diameter in the range of 50 nm \cite{Ganser} and a length in the range of 100 nm. The confinement free energy cost of a 10 kb RNA molecule inside such a capsid would be estimated to be quite small. Actually, the size of the ss RNA genome molecules of HIV and other retroviruses is reduced further because they are \textit{condensed} by the NCp7 fragments. The RNA genome molecules of HIV are believed to occupy only about 20 \% of the average volume of a mature HIV capsid. Confinement free energies would thus be expected to be insignificant for HIV.

In this paper we propose that confinement forces \textit{do} play an important role in the HIV life cycle. The conical HIV capsid contains a copy of the protein \textit{Reverse Transcriptase} (RT). The RT enzyme converts ss RNA molecules into ds DNA molecules. Until recently it was commonly accepted that reverse transcription (RTion) in retroviruses, including HIV, takes place within the cytoplasm of the infected cell after the uncoating of the mature capsid \cite{Mirambeau}. Recent studies argue that the inner capsid appears to remain intact during most of the RTion \footnote{\textit{Retroviruses}. Edited by Coffin JM, Hughes SH, Varmus HE, Cold Spring Harbor (NY); Cold Spring Harbor Laboratory Press (1997)}, and that RTion stalling significantly slows down uncoating \cite{Hulme}. It indeed would seem clear that the RTion process should be much more efficient if it took place inside a small ``reaction vessel" where the concentration of reactants and enzymes would be much higher than in the cytoplasm of the host cell. This suggests that RTion leads to uncoating. The transformation of an ss RNA genome molecule into a ds DNA molecule indeed transforms the very flexible annealed branched secondary structure of ss RNA into a linear B-DNA polymer with a substantial bending stiffness (the persistence length of B-DNA is in the range of 50 nm). If the HIV capsid is sufficiently fragile, then the increased DNA confinement energy could potentially lead to uncoating. The forces would of course be much weaker than for the case of phages, which means that retroviral capsids should be extremely fragile. However, the NCp7 fragments are likely to be condensing agents for ds DNA genome molecules as well. It is known that the presence of millimolar concentrations of condensing agents induces long B-DNA molecules to spontaneously condense into tightly wound toroids with a diameter in the 100 nm range \cite{Hud}. In fact, provided the mature capsid is intact during the RTion, the NC concentration inside the capsid is in the milliMolar range, and should be sufficient to completely condense the dsDNA genome. This evidently would weaken the confinement forces exerted by the DNA. If the strength of the condensing agents is properly matched with the fracture strength of the capsid then this could produce uncoating - during RTion - for a reasonably well-defined B-DNA length. Increasing the strength of the condensing agents could shrink the DNA to the extent that there is no uncoating while decreasing the strength could produce premature uncoating. Recent findings showing that there is an optimum stability for the mature HIV capsid that is critical for viral infectivity \cite{Shah} would be in support of such a proposal. 

Physical descriptions that were developed to describe the very large DNA confinement forces inside phages \cite{Kindt} cannot be directly applied to the case of retroviruses. It is the aim of this article to provide a physical description of the uncoating of retroviruses by RTion, using only basic soft matter physics, and to find the proper matching condition between the strength of dsDNA condensing agents and the fracture strength of the capsid. In the next two sections we first provide simple descriptions of the mechanical properties of DNA toroids and of viral shells. In the last section we construct an ``uncoating diagram" that could be used to interpret experiments on HIV capsid uncoating.

\section{Mechanical Properties of a DNA toroid}

\begin{figure}
\centering
\includegraphics[width=0.49\textwidth]{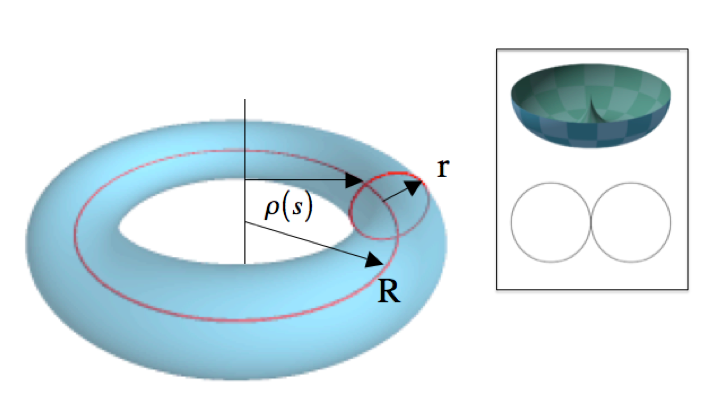}\\
\caption{\label{fig:toroid} Geometry of a ``skinny" DNA toroid. The major axis is R is significantly larger than the minor axis is r. The curvature radius of a DNA strand at an arc-distance s is indicated by $\rho(s)$. The inset case shows the``fat" toroid: the limiting case R=r. (credit: Wikipedia)}.
\end{figure}

Assume a B-DNA filament of $N$ bps with a contour length $L=Nb$. Here, $b\approx0.34 nm$ is the length of one B-DNA bp along the filament axis. During RTion, the number $N$ increases from zero to a maximum value $N_{HIV}\sim 10^4$.  The DNA filament is self-attracting, due to the presence of condensing agents - the NC proteins -,  which causes it to be wound into the shape of toroidal spool. Figure \ref{fig:toroid} shows the geometry of a toroid. The major radius will be denoted by $R$ and the minor radius by $r$.  The spacing between the filaments in the toroid will be denoted by $d$ $(\approx$3 nm).
By equating the volume of a cylinder of length $L$ and diameter $d$ to the volume of the toroid times the occupied volume fraction of a hexagonally close packed bundle of cylinders, it follows that the minor and major toroid radii of the toroid are related by $r\sim d(\frac{L}{R})^{1/2}$ (we will drop all numerical constants and prefactors in the following).

The free energy of the DNA toroid will be written as a cohesive contribution $F_{c}$ due to the self-attraction  plus a bending energy cost $F_{b}$. Energies will be expressed in units of the thermal energy $k_B T$. The cohesive free energy $\alpha$ per bp depends on the nature of the condensing agent. It has not been measured for HIV NCp7 but it is in the range $0.01-0.5$ for conventional condensing agents. The cohesive free energy of the toroid is then
\begin{equation}\label{399}
F_c \sim -(\alpha/b )\bigg(L-\frac{rR}{d}\bigg).
\end{equation}
The second term inside the bracket is an estimate of the loss of cohesion of bps located on the surface of the toroid. The bending energy of the toroid is estimated by treating dsDNA as a semi-flexible polymer of length $L$: 
\begin{equation}\label{399}
F_b\approx\frac{l_p}{2}\int_{0}^{L}ds\frac{1}{\rho^2(s)}.
\end{equation}
Here, $\rho(s)$ is the local curvature radius at an arclength $s$ along the DNA filament, approximated as the distance to the central axis of the toroid ($l_p$ is the persistence length). Using the fact that $dS/d^2$ is proportional to the number of times the DNA chain pierces an area element dS, the line integral along the DNA molecule can be converted into a surface integral over a cross-section $A$ of the toroid: 
\begin{equation}\label{399}
F_b\sim\frac{ l_p}{d^2}\int_{A}\frac{dS}{\rho(\vec{r})}.
\end{equation}
Using a polar coordinate system measured from the center of a cross-section, the integral can be carried out to give
\begin{equation}\label{399}
F_b(R)\sim\frac{ l_p R}{d^2}\bigg(1-\sqrt{1-\bigg(\frac{R_{min}}{R}\bigg)^3}\bigg).
\end{equation}
The singularity of the bending free energy at $R_{min}$ is the point where the minor radius $\rho$ of the toroid equals the major radius $R$ so where $R_{min}\sim d(\frac{L}{R_{min}})^{1/2}$. This gives
\begin{equation}\label{399}
R_{min}(L)\sim(d^2L)^{1/3}.
\end{equation}
(the surface of this toroid is known as a ``horn torus"). Adding the bending and cohesive terms gives a variational free energy for the radius $R$ of the toroid

\begin{equation}\label{399}
F(R)\sim\frac{ l_p R}{d^2}\bigg(1-\sqrt{1-\bigg(\frac{R_{min}}{R}\bigg)^3}\bigg)+(\alpha/b)(LR)^{1/2}
\end{equation}
where we dropped terms independent of $R$.

 \subsection{Skinny toroid}

If the major radius R of the toroid is large compared to the minor radius $r$, so if $R  >>  R_{min}$, then one can expand the square root in powers of $R_{min}/R$. To lowest order

\begin{equation}\label{Soft}
F_t(R)\sim\frac{ l_p L}{R^2}+(\alpha/b){(RL)^{1/2}}.
\end{equation}
This free energy is minimized by a major radius 

\begin{equation}\label{Par}
R^*(L)\sim (bl_p/\alpha)^{2/5}L^{1/5}
\end{equation}
that depends weakly on the DNA length. The condition $R^*(L) >> R_{min}$ holds for DNA lengths less than $L^*$ with

\begin{equation} 
L^*\sim\left(bl_p/\alpha\right)^3/d^5.
\end{equation} 

The equilibrium elastic deformability of the toroid is measured by the \textit{stiffness} $K_t\equiv\frac{\partial^2 F}{\partial R^2}$ evaluated at $R=R^*(L)$. This stiffness plays the role of a spring constant for deformation of the toroid. Using Eq.~\ref{Soft}, one finds

\begin{equation}\label{Comp3}
K_t(L)/k_BT\sim l_p(\alpha/bl_p)^{8/5}L^{1/5}.
\end{equation}
Like the major radius, the stiffness of a skinny toroid is nearly independent of DNA length.  
The deformation free energy $\Delta F_t(R)$ of a skinny toroid can be expressed as

\begin{equation}\label{Comp3}
\Delta F_t(R)\sim (1/2)K_t(L)\left(R-R^*(L)\right)^2+......
\end{equation}

 \subsection{Fat toroid}

For $L >> L^*$, the toroid swells up and the central hole closes up. The minor and major radii both become comparable to $R_{min}$. Expand $F(R)$ in powers of $x=(R-R_{min})/R_{min}$. Define $\delta f(x) = \frac{F(R)-F(R_{min})}{F(R_{min}}$, where $F(R_{min})\sim \frac{l_p R_{min}}{d^2}$ is of the order of $10^2$. Using Eq.\ref{399}

\begin{equation}\label{F(x)}
\delta f(x)\sim -x^{1/2}+\frac{\alpha d^2 }{bl_p}\left(\frac{L}{R_{min}}\right)^{1/2}x.
\end{equation}
 Minimization gives $x\sim\left(\frac{bl_p}{\alpha d^2}\right)^2\left(\frac{R_{min}}{L}\right)$, which vanishes in the limit of large $L$. The second derivative of the free energy with respect to $R$ depends on $x$ as $1/x^{3/2}$, which diverges in the limit of small $x$. That means that the stiffness of a fat toroid diverges with genome length as $L^{2/3}$ for $L>>L^*$. Matching to the skinny toroid at $L\sim L^*$ gives the scaling relation:

\begin{equation}\label{F(x)}
K_t(L)/K_T(L^*)\sim  (L/L^*)^{2/3}
\end{equation}
where $K_t(L^*)/k_BT\sim \frac{\alpha}{db}\sim\alpha$/nm$^2$ is obtained by matching to the skinny toroid at $L=L^*$.

\section{Mechanical Properties of a Viral Shell}
\label{Mech}

\begin{figure}
\centering
\includegraphics[width=0.2\textwidth]{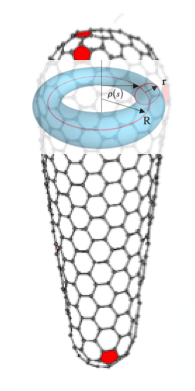}\\
\caption{\label{Figure3.png} Conical HIV capsid composed of hexamers and twelve pentamers (red). From Ref.~\cite{Ganser}. A toroid is fitted at the maximum cross section. The sum of the major and minor radii equals the shell radius.}
\end{figure}

As a DNA toroid grows in size and its diameter approaches that of the capsid, it will start to exert a force on the capsid. We will restrict ourselves in the following to spherical capsids. Though there are spherical retroviruses (such as M-MuLV), the HIV capsid is conical, as already noted. If one assumes however that the DNA toroid is placed in the plane of the maximum cross-section, as shown in Fig.\ref{Figure3.png}, then the deformation and fracture of the conical capsid is expected be similar to that of a spherical capsid with the radius of the hemisphere.   
The force exerted by the growing toroid on the capsid, which is localized along the equator of the capsid and which cannot be described in terms of a uniform pressure, will be described using continuum elasticity theory. The mechanical properties of plant and phage viral capsids have been extensively studied, both experimentally and theoretically \cite{Roos} and the theory of thin elastic shells was found to provide a practical theoretical framework to describe the mechanical properties of such shells \cite{Lidmar}. A thin elastic shell is characterized by two moduli. First, the \textit{Bending Modulus} $\kappa$ describes the energy cost of altering the out-of plane-curvature of the shell. Next, the 2D \textit{Young's Modulus} $Y$ describes the energy cost of in-plane elastic deformation of the shell, such as stretching. The dimensionless combination $\gamma=4\pi {R_c}^2 Y/\kappa$ is known as the ``F\"{o}ppl-von K\'{a}rm\'{a}n Number". Both moduli have been measured for bacteriophage and plant viral shells by micromechanical studies \footnote { $\kappa$ is in the range of $10 - 100\thinspace k_BT$ and $Y$ is in the range of GPa times the shell thickness of about one nm \cite{Roos}.} but they are not known for retroviruses.

The elastic free energy of a thin shell is now

\begin{equation}
E_c=\frac{1}{2}\int dS \left(\kappa H^2 +2\mu \epsilon_{ij}^2+\lambda\epsilon_{ii}^2\right)
\end{equation}
The first term is the bending energy with $H=1/R_1+1/R_2$ the \textit{mean curvature} where $R_1$ and $R_2$ are the principal radii of curvature at a given point of the surface. The second term is the stretching eneregy with $\lambda$ and $\mu$ the Lame constants, related to the Young's Modulus by $Y=\frac{4\mu(\mu+\lambda)}{2\mu+\lambda}$.  If a thin elastic shell with radius $R_c$ is indented by a localized radial point force over a distance $\zeta$, then this creates a circular ``dimple". Let $\delta$ be the radius of the dimple. The induced curvature of the dimple is of the order of $\zeta/\delta^2$ so the bending energy is of the order $\kappa (\zeta/\delta)^2$. The induced stretching strain due to the local increase in area is of the order$\zeta/R_c$  so the stretching energy is of the order of $Y (\zeta\delta/R_c)^2$. Minimizing the sum of bending and stretching energy with respect to $\delta$ leads to $\delta\sim\left(\frac{\kappa{R_c}^2}{Y}\right)^{1/4}\sim R_c/\gamma^{1/4}$, a well-known result \footnote{Landau and Lifshitz, \textit{Theory of Elasticity}, Ch.15. For a retroviral shell, the dimple radius would be in the range of 10 nm if the HIV elastic moduli would be similar to that of phages.}. The deformation energy is harmonic in $\zeta$ and of the form $E(\zeta)\sim\frac{(Y\kappa)^{1/2}}{R_c}\zeta^2$ of the indentation involves both bending and stretching \footnote{Spring constants for plant and phage viral shells are in the range of 0.5 N/m. For the larger retroviral shells the spring constant would be in the range of 0.1 N/m if they would have elastic properties similar to that of phages.}. If one replaces the point force by a \textit{line force} exerted along an equator of a spherical capsid, then the deformation energy cost could be estimated as that of a sequence of $R_c/\delta$ dimples. The ``equatorial" deformation energy $E_{eq}$ is estimated as $E(\zeta)$ times $R_c/\delta$ or 
 
\begin{equation}
E_{eq}(\zeta)\sim\frac{Y^{3/4}\kappa^{1/4}}{{R_c}^{1/2}} \zeta^2
\end{equation}
One could can define an equatorial stiffness $K_c\sim Y^{3/4}\kappa^{1/4}/{R_c}^{1/2}$

We conclude this section by recalling that the fracture strength of a material is determined by a materials parameter known as the ``strain-at-failure",  which we will denote by $\epsilon_c$. The failure stress, or yield stress, is then of order $Y\epsilon_c$. For strains less than $\epsilon_c$, deformations are reversible and the stress-strain relation is linear. So called ``brittle" materials break when the strain exceeds $\epsilon_c$, while ``plastic" materials undergo irreversible, plastic deformation before fracture. Though plastic deformation has been observed for viral shells \cite{Roos}, we will focus here on the simpler case of brittle fracture with $\epsilon_c$ small compared to one.

\section{Uncoating Conditions and Uncoating Diagram for Brittle Capsids}

We now apply the results of the last two sections to construct an ``uncoating diagram" that can be used as a guide to predict if and how a DNA toroid growing inside a spherical viral shell of radius $R_c$ will cause the shell to fracture or fall apart. Values for genome volume fractions given below are all with respect to this spherical volume rather than the full volume of a conical capsid.

\begin{figure}
\centering
\includegraphics[width=0.50\textwidth]{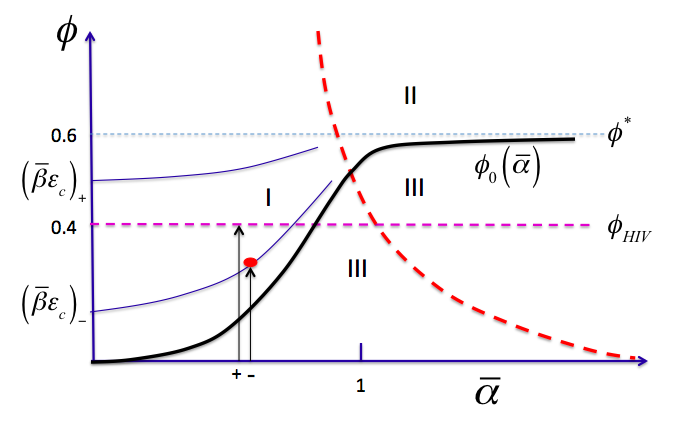}\\
\caption{\label{Diagram} Uncoating diagram. The vertical axis $\phi$ is the ratio of the volume occupied by B-DNA over the volume of the (undeformed) spherical caspid. The estimated volume fraction $\phi_{HIV}\sim0.4$ of fully converted B-DNA is shown as a purple dashed line. The horizontal axis $\bar{\alpha}$ is the cohesion energy per base pair in dimensionless units. The thick black line $\phi_0(\bar\alpha)$ is the condition for the DNA toroid to have the same radius as the capsid. In the limit of large $\bar\alpha$, $\phi_0(\bar\alpha)$ approaches $\phi^*\approx0.6$, the volume fracion of a horn torus inscribed inside a sphere. It is also the volume fraction for the uncoating of capsids in the limit of the critical strain going to zero. In region III below the black line, uncoating by confinement pressure is not possible.
The dashed red line indicates cross over from skinny to fat toroids. Uncoating taking place on the left side of the red line involves skinny toroids, and fat toroids on the right. The cohesion energy $\bar\alpha$ is normalized so $\bar\alpha=1$ corresponds to the intersection of $\phi_0(\bar\alpha)$ with the red dashed cross over line. In region I on the left of the red dashed line, uncoating is driven by skinny toroids and in region II by fat toroids. The two thin black lines marked + and - are examples of uncoating curves $\phi(\bar\alpha)$ for two different values of the materials parameter $\bar\beta\epsilon_c$, that determines uncoating of brittle capsids. Uncoating can be visualized as taking place for a fixed value of $\bar\alpha$ with $\phi$ increasing from 0 to $\phi_{HIV}$ as indicated by the two vertical arrows. For the more robust capsid (+) with the larger value $(\beta\epsilon_c)_+$, $\phi(\bar\alpha)$ lies above $\phi_{HIV}$. There can be no uncoating by RTion.  For the more fragile capsid (-), with $(\beta\epsilon_c)_-$ well below $\phi_{HIV}$, the uncoating line reduces to the black curve  Uncoating is indicated by the red dot. 
}
\end{figure}

The uncoating diagram shown in Fig.\ref{Diagram} is a central result of this paper. 

 $\phi\sim\bar\alpha^2$ $\phi\sim1/\bar\alpha^3$ $\phi(\bar\alpha=0)\sim\bar\beta\epsilon_c$, $\phi\sim\bar\alpha^2$.It follows that if $\bar\alpha$ exceeds a critical value of the order of  $(\phi_{HIV})^{1/2}$ then there can be no uncoating because the toroid is too small. 
The vertical axis is the \textit{volume fraction} $\phi$, i.e., the ratio of the volume of the DNA and that of the undeformed spherical capsid. It can also be expressed as the total length of the DNA molecule normalized by the length of a DNA molecule that would completely fill the spherical capsid so $\phi(L)\sim Ld^2/R^3$. The volume fraction $\phi^*=3\pi/16\sim 0.59$ indicated in the diagram is that of a horn torus with $r=R=R_c/2$ inscribed inside the capsid. The dashed line $\phi_{HIV}\sim0.4$ is the estimated maximum DNA volume fraction of HIV, which is thus less than $\phi^*$. The horizontal axis is the normalized cohesive energy $\bar{\alpha}=\alpha/\alpha_{0}$ per base pair of the toroid with$\alpha_{0}=\frac{bl_p}{dR_c}\sim0.2$. Finally, we ialso ntroduce a dimensionless capsid stiffness $\bar\beta$ defined by $K_t=\bar\beta K_{cr}$ where $K_{cr}=\frac{k_BT l_p}{R_cd^2}$ has the dimensions of a spring constant.

Various relationships simplify in these units. The relation for the cross over DNA length between skinny and fat toroids $L^*\sim\left(bl_p/\alpha\right)^3/d^5$ reduces to $\phi\sim 1/\bar{\alpha}^3$, shown as the dashed heavy red line in the diagram. Next, the condition $R_c\sim L^{1/5}(bl_p/\alpha)^{2/5}$ for a skinny toroid to touch the capsid wall (obtained from Eq.\ref{Par}) reduces to $\phi\sim\bar{\alpha}^2$. The heavy black line $\phi_0(\bar\bar\alpha)$  in Fig.\ref{Diagram} is the volume fraction at which the toroid has the same diameter as the capsid for a given value of $\bar\bar\alpha$. In the limit $\bar\beta\epsilon_cF= 0$, so for extremely fragile capsids, it also is the uncoating volume fraction. For small $\bar\bar\alpha$, $\phi_0(\bar\bar\alpha)\sim\bar\alpha^2$ while $\phi_0(\bar\bar\alpha)\sim\phi^*$ for large $\bar\bar\alpha$. Uncoating is not possible in the region of the diagram marked III below $\phi_0(\bar\bar\alpha)$. The intersection of $\phi_0(\bar\bar\alpha)$ with the skinny-to-fat-toroid transition line is near $\bar\alpha=1$. From the fact that $\phi_{HIV}\sim0.4$ is less than $\phi^*=3\pi/16\sim 0.59$, we deduce (i) that capsid uncoating by RTion for HIV-1 is possible only for values of $\alpha$ less than $\alpha_{cr}\sim0.2$ and (ii) that uncoating must take place for skinny toroids. 

We now express, in these units, the strength of the self attraction $\bar\alpha(\phi)$ such that uncoating takes place at a certain volume fraction (presumably of the order of $\phi_{HIV}$). In the previous section, we noted that according to elasticity theory the fracture condition for the capsid is that the strain $\epsilon\sim(R-R_c)/R_c$ had to exceed the critical value $\epsilon_c$ with $R$ the maximum radius of the capsid as it is deformed by the toroid.  We determine $R$ by minimizing the sum of the capsid and toroid free energies:

\begin{equation}
F(R) \sim\frac{ l_p L}{R^2}+(\alpha/b){(RL)^{1/2}}+\frac{1}{2}K_c(R-R_c)^2.
\end{equation}
Setting the derivative of $F(R)$ to zero gives.

\begin{equation}\label{force}
\frac{ l_p L}{R^3}-(\alpha/b){(L/R)^{1/2}}\sim K_c(R-R_c)
\end{equation}
If the critical strain of the HIV capsid is significantly less than one - the limit of brittle capsids - then we can equate $R$  to $R_c$ on the left hand side. On the right hand side, we equate $R-R_c$ to $\epsilon_cR_c$. In the new units, this condition reduces to:

\begin{equation}\label{strain}
\bar\beta\epsilon_c\sim {\phi}-\bar\alpha{\phi^{1/2}}
\end{equation}
The desired relation is then

\begin{equation}\label{ALPHA}
\boxed{\bar\alpha(\phi)\simeq\frac{c_1\phi-c_2\bar\beta\epsilon_c}{\phi^{1/2}}}
\end{equation}
with $c_{1,2}$ numerical constants. Note that $\bar\alpha(\phi)$ only depends on the mechanical properties of the capsid through the dimenionless parameter $\bar\beta\epsilon_c$, which has the physical meaning of a dimensionless fracture stress. 

We can make a number of conclusions from this relation. First, since $\bar\alpha(\phi)$ monotonically increases with $\phi$, the uncoating condition can never be satisfied if $\bar\alpha$ exceeds $c_1\phi_{HIV}^1/2$. Physically, this means that the maximum equilibrium size of the toroid should exceed that of the capsid. In standard units, 

\begin{equation}\label{con1}
\alpha <c_1\frac{bl_p}{dR_c}({\phi_{HIV}})^{1/2}
\end{equation}
 
Next, since $\bar\alpha(\phi)$ should be positive, a necessary condition for HIV uncoating by RTion is that  $\bar\beta\epsilon_c$ should be less than $(c_1/c_2)\phi_{HIV}$. I the DNA volume fraction inside the capsid is smaller than $(c_2/c_1)\bar\beta\epsilon_c$, it would not be able to uncoat that capsid - even if there was no DNA condensation - because the stress $\bar\beta\epsilon_c$ exerted by B-DNA would be too weak. In standard units, this simple condition turns into a rather more complex condition:

\begin{equation}\label{con2}
\epsilon_c\left(\frac{c_2Y^{3/4}\kappa^{1/4}{R_c}^{1/2}d^2}{c_1{k_BT l_p}}\right)<\phi_{HIV}
\end{equation}

One can also invert Eq.\ref{ALPHA}, to obtain the volume fraction  $\phi(\bar\alpha)$ at which uncoating takes place for a given $\bar\alpha$.  The limiting properties of $\phi(\bar\alpha)$ are

\begin{equation}\label{VWSA}
 \phi(\bar\alpha\to0)\sim\bar\beta\epsilon_c+\bar\alpha{(\bar\beta\epsilon_c)^{1/2}}+..
\end{equation}

while for large $\bar\alpha$:

\begin{equation}\label{VSSA}
 \phi(\bar\alpha\to\infty)\sim\bar\alpha^2+..
\end{equation}

The uncoating diagram Fig.~\ref{Diagram} shows two different examples of $\phi(\bar\alpha)$, one with the fracture stress $\bar\beta\epsilon_c$ less than $\phi_{HIV}$ marked by a minus sign and one with $\bar\beta\epsilon_c$ larger than $\phi_{HIV}$ marked by a + sign. The progression of RTion inside the capsid can be visualized as a time-dependent volume fraction $\phi(t)$ of dsDNA inside the capsid that increases over time from $\phi=0$ to $\phi_{HIV}~0.4$, at some fixed value of the DNA self-attraction $\bar\alpha$. The two vertical arrows are examples of two systems with the same value of $\bar\alpha$ but different fracture stresses (marked + and -). The + case would be an example of a capsid that is so robust that it cannot be uncoated by RTion of HIV genome while the - case is an example of a capsid that is sufficiently fragile so it can be uncoated by RTion beyond a critical volume fraction marked by the red dot.

\section{Conclusion}

The uncoating diagram Fig~\ref{Diagram}, Eq.~\ref{ALPHA} and the two conditions on the fracture stress and the strength of the selff-attraction (Eqs.~\ref{con1} and~\ref{con2}) are our main results. How do they apply to HIV?. According to Eq.\ref{con1}, the dimensionless stress $\beta\epsilon_c$ at uncoating should be less than $\phi_{HIV}$ times a number of the order of one. If the materials properties of HIV capsids - in terms of $\kappa$, $Y$, and $\epsilon_c$ - would be similar to those of the capsids of small RNA viruses (such as CCMV) then $\beta\epsilon_c$ would be of the order of $10^2$, so either the capsid stiffness $K_c$ or the fracture strain $\epsilon_c$ should be much less for HIV or both.

According to the diagram, $\bar\alpha$ must be less than a critical value of about 0.6, which is where the dashed purple line intersects the black line. Using $\alpha_{cr}\sim\frac{bl_p}{dR_c}$, we find that $\alpha$ must be less than $\approx0.15$. The DNA toroid is necessarily skinny before and at uncoating. In other words, for the proviral dsDNA to uncoat the capsid at such low volume occupancy, the toroidal globule should have a large internal whole, and an external diameter larger then the one of the capsid . The later is only possible if the DNA self-attraction is sufficiently weak, such that the DNA toroid is only loosely wound. The uncoating line Eq~\ref{UN} must run well to the left of the red curve. It follows that $\bar{\beta}$ should be significantly below one. Also, it follows that the mature HIV capsid should be rather fragile, such that the product of its rigidity and critical strain b*ec are less then FiHIV~0.4. If the critical strain ec of the HIV capsid (still to be measured) is close to one, i.e. the mature HIV capsid is as elastic as typical shells of the small plant viruses, it implies that its rigidity parameter b should be <fiHIV~0.4. The later can be expliciity related to the bending and stretching capsid rigidities (still to be measured), and further related to the strength of protein-protein interactions that form the capsid. Mutations of the capsid proteins leading to the stronger or weaker capsids were both shown to be detrimental to the HIV live cycle (Aiken), consistent with our prediction that the more stable capsid will never get uncoated by RTion, while the weaker capsid will uncoat pre-maturely before the RTion transition completion, thereby leading to the loss of essential proteins and RTion termination.It follows that the capsid should be fragile. In summary, it follows from our theory that the uncoating of HIV by RTion requires fragile capsids, relatively weak condensing agents, and skinny DNA toroids. 
For the capsid elasticity constant Kc this implies that it should be  $Kc<=kBT*lp/(Rc*d^2)~0.001N/m$, i.e. about 100-fold less than the elasticity of the small plant viruses measured by micromanipulation. While the elasticity of mature HIV capsid was not yet measured, this prediction is consistent with the very low capsid stability, that had so far precluded its biochemical purification and its mechanical studies.

Just as the capsid stability should be low and tuned for uncoating by RTion, the strength of NC-induced DNA self-attraction appears to be also  optimized for that cause. The 0.15 KbT of attractive free energy per bp estimated above is a reasonable value, when compared to the analogous

The estimated maximum value for the cohesion energy is a reasonable value for the NC-induced DNA self-attraction free energy, based on analogous values for the DNA self-attraction with simple non-biological cations such as CoHex$^{3+}$ or Spd$^{3+}$, that have an effective charge similar to that of HIV NC (about 3.5+) \cite{Nguyen}. Consistent with these ideas, there is evidence that mutations in NCp7 that reduce its condensation ability (e.g., by reducing the positive charge) lead to the major RTion defects and complete loss of viral infectivity \cite{Lidmar}\footnote{Distinct nucleic acid interaction properties of HIV-1 nucleocapsid protein precursor NCp15 explain reduced viral infectivity
Wang, Wei; Naiyer, Nada; Mitra, Mithun; Li, Jialin; Williams, Mark C; Rouzina, Ioulia; Gorelick, Robert; Wu, Justin; Musier-Forsyth, Karin
NAR-00038-2014 - submitted}.In a future publication, we plan to discuss the case of capsid uncoating by fat DNA toroids (Region II in the uncoating diagram) as well as implications of this uncoating diagram for the HIV life cycle .

Acknowledgements: IR acknowledges support from NIH Grant GM065056 and Professors Karin Musier-Forsyth, Alan Rein and Judith Levine for support and discussions. RFB acknowledges support by the Institute of Advanced Studies of the TUM and by the NSF under DMR Grant 1006128. RB thanks S. Grosberg for instruction in the use of the different symbolic representations of the approximate sign.



\end{document}